\documentstyle[prl,aps,psfig,epsf]{revtex}

\begin{document}
\draft
\title{Comment on: ''Radiation-Induced ''Zero-Resistance State'' and the Photon
Assisted Transport''}
\author{A.F. Volkov$^{1,2}$}
\address{$^{1}$ Theoretische Physik III,\\
Ruhr-Universit\"{a}t Bochum, D-44780 Bochum, Germany\\
$^{2}$Institute of Radioengineering and Electronics of the Russian \\
Academy of Sciencies, Moscow 103907, Russia.}
\date{\today}
\maketitle

\begin{abstract}
Shi and Xie [8] predicted the N-shape current-voltage characteristic $(CVC)$
for a 2DEG in a magnetic field where the zero-resistance state has been
observed in recent experiments. However it is known that in the absence of a
magnetic field the zero resistance state (zero differential resistance
state) is achieved in a system with the S-shape $CVC$. The difference in the
behaviour of systems with N- and S-shape $CVC$ was studied more than three
decades ago (see for example the review [9] and references therein) and is
briefly explained in this Comment. At present it is not clear whether the
N-shape $CVC$ may lead to the zero resistance state in a 2DEG in a magnetic
field.
\end{abstract}

An interesting effect has been observed in recent papers \cite
{Klitzing,Zudov}: the resistance of a 2-DEG subjected to microwave
irradiation drops to zero in some interval of an applied magnetic field $B.$
Possible mechanisms for this phenomenon are discussed in a number of papers 
\cite{Philips,Yale,Andreev,Anderson,Shrivastava,Shi,Raikh,Mikhailov}. It was
shown in Refs. \cite{Yale,Shi,Raikh} that in the presence of irradiation and
a sufficiently strong magnetic field the conductivity $\sigma _{xx}$ may
become negative in a weak electric field $E_{x}$. In addition, using a
simple model, the authors of Ref. \cite{Shi} have calculated the
current-voltage characteristic $I(V)$ of an irradiated system with a
density-of-states periodic in energy $\varepsilon $. They have demonstrated
that not only regions with negative conductance $G$, but also regions with
positive $G$, may have negative differential conductance $G_{d}=dI/dV$ on
the current-voltage characteristic ($CVC$) curve.

In the present Comment I would like to note that effects in systems with
negative differential conductance (or resistance) depend in a crucial way on
the type of $CVC$, i.e. whether it has the N- or S-shape form (see, for
example, the review \cite{KoganVolkovUs} and references therein). It was
established long ago that the states corresponding to regions with negative $%
G_{d}$ on $CVC$ are unstable with respect to nonhomogeneous fluctuations.
However the type of instability and final nonhomogeneous state which is
formed as a result of the instability depend on the type of $CVC.$

In the case of the N-shape $CVC$ (three voltages $V_{i}$ correspond to one
current $I$) a domain with a strong electric field (as happens, for example,
in the Gunn effect) arises as a result of an instability in a homogeneous
initial state (it is assumed that the total voltage is fixed). This domain
moves in the direction of the applied electric field $E=V/L$ ($L$ is the
length of the sample). In the presence of the high field domain the $CVC$
changes drastically: an almost flat region (plateau) appears on the $I(V)$
curve. The differential conductance $G_{d}$ (but not the resistance $%
R_{d}=dV/dI$) is zero on this plateau. The current $I_{pl}$ corresponding to
this flat part of $CVC$ depends on the particular form of the $CVC$. In the
case of a ''symmetric'' $I(V)$ curve with absolute negative conductance
(i.e. min\{$dI/dV$\} corresponds to $V=0$ ) this current is equal to zero
(that is the total conductance is zero). This means that when the bias
voltage varies in some limits, the current remains equal to zero. The $CVC$
obtained in Ref. \cite{Shi} on the basis of a toy model can be regarded as a
chain of N-shape $I(V)$ curves.

In the case of the S-shape $CVC$ (three currents $I_{i}$ correspond to one
voltage $V$) small perturbations increase in the direction transverse to the
current $I$ if this current corresponds to the part of $CVC$ with negative $%
R_{d}$. As a result of this instability, a domain of a strong current
density arises in the sample (it is assumed that total current is fixed,
otherwise the system goes over to a state on the stable part of the $CVC$).
In the presence of this current domain (or a filament with a higher current
density) the $CVC$ is modified so that an almost vertical part appears on
the $I(V)$ curve ($R_{d}$ is close to zero). The electric field $E_{ver}$
corresponding to this part of the $CVC$ is again determined by the shape of $%
CVC$. All these statements concerning the S-type of $CVC$ are based on the
study of the so called superheating mechanism of the S-shape $I(V)$ curve 
\cite{KoganVolkovZh} . In the case of a ''symmetric'' S-shape $CVC$ with
negative $G$ (i.e. min\{$dV/dI$\} corresponds to $I=0$) the field $E_{ver}$
is zero (when the bias current varies in some limits, the voltage across the
sample is zero ). Similar conclusions were made in a recent paper on a
simple phenomenological model \cite{Andreev}.

Of course the behaviour of a system with the N- or S-shape $CVC$ in the
absence or  presence of a magnetic field $B$ is different. In the latter
case the Hall field $E_{y}$ arises for example if the transverse current $%
I_{y}$ is zero. The form of the $CVC$ of a homogeneous 2DEG (in the absence
of domains of the electric field or current) depends on whether the Hall
current $I_{y}$ or the Hall field $E_{y}$ is zero. For instance, if the $%
I_{x}(V_{x})$ dependence corresponds to the N-shape $CVC$ in the absence of
the Hall field $E_{y}$ $(I_{y}\neq 0)$, this correspondence changes
completely in the absence of the Hall current $I_{y}$ ($E_{y}\neq 0$). It
acquires a complicate form and can not be assigned to the N- or S-shape type
of $CVC$ (several values of $I_{x}$ correspond to one $E_{x}$ and several
values of $E_{x}$ correspond to one $I_{x}$). Therefore one has to analyse
the type of instability and the form of the final state of the system
separately for these two cases ($I_{y}=0$ or $E_{y}=0$).

To summarize, one can conclude that the absolute negative conductance
obtained in Refs. \cite{Yale,Shi,Raikh} is not sufficient to explain the
observed zero resistance (or differential resistance) states. In a system
with negative conductance at low dc fields $E,$ the $CVC$ may have either
the N- or S-shape form, and the behaviour of the system will be different in
these both cases. I also would like draw attention to the fact that negative
conductance was predicted long ago in Refs.\cite
{Ryzhii,Suris,Elesin,V'yurkov} (2DEG and 3DEG in a quantizing magnetic field
in the presence of microwave irradiation) and in Refs.\cite{Epstein}
(superlattices under ac irradiation).

I would like to thank Sh.M.Kogan as well as B.Huckestein, N.Rick and J.Shi
for useful discussions and comments.

\end{document}